\documentclass[twocolumn, english]{article}
\usepackage[sorting = none, backend = biber, style=numeric-comp]{biblatex}
\usepackage{graphicx}
\DeclareUnicodeCharacter{0301}{\'{e}}
\usepackage{hyperref}
\usepackage{authblk}
\usepackage{geometry}
\usepackage{adjustbox}
\usepackage{svg}
\usepackage{amsmath}

\usepackage{color}
\usepackage{soul}
\soulregister\cite{7} 
\soulregister\ref{7} 
\soulregister\eqref{7} 

\begin{document}

\title{Spectral Shaping in a Multimode Fiber by All-fiber Modulation}

\author{Zohar Finkelstein}
\author{Kfir Sulimany}
\author{Shachar Resisi}
\author[1,*]{Yaron Bromberg}

\affil{Racah Institute of Physics, The Hebrew University of Jerusalem, Jerusalem 91904, Israel}
\affil[*]{Corresponding author: yaron.bromberg@mail.huji.ac.il}

\date{}

\twocolumn[\begin{@twocolumnfalse}
\maketitle
\begin{abstract}
In the past few years, there is a renewed interest in using multimode fibers for a wide range of technologies such as communication, imaging and spectroscopy. Practical implementations of multimode fiber in such applications, however, are held back, due to the challenges in dealing with modal dispersion, mode coupling and the fiber’s sensitivity to mechanical perturbations. Here, we utilize these features of multimode fibers to generate all-fiber reconfigurable spectral filters. By applying computer-controlled mechanical deformations to the fiber, along with an optimization algorithm, we manipulate the light propagation in the fiber and control its output field. Using this approach we demonstrate tunable bandpass filters and dual-band filters, with spectral resolutions as low as 5pm.\newline 

\end{abstract}

\end{@twocolumnfalse}]


\section{Introduction}

In the past decade there is ongoing effort to utilize multimode optical fibers (MMF) in a wide range of photonic applications such as optical fiber communication \cite{Richardson2013, Ploschner2015,puttnam2021space}, high energy fiber lasers \cite{wright2017spatiotemporal,teugin2019spatiotemporal,haig2022multimode}, quantum optics\cite{defienne2016two,valencia2020unscrambling,leedumrongwatthanakun2020programmable,cao2020distribution,sulimany2022all}, nonlinear optics \cite{mafi2012pulse,wright2015controllable, krupa2017spatial, tzang2018adaptive,krupa2019multimode, teugin2020controlling,pourbeyram2022direct} and imaging \cite{vcivzmar2012exploiting,choi2012scanner,gu2015design,caravaca2017single,borhani2018learning,leite2021observing,resisi2021image,lee2022confocal}. The challenge of MMF-based technologies is that transverse modes of the fiber are interrupted by inter-modal interference, mode coupling, and modal dispersion. Therefore, the information delivered by a MMF is scrambled, forming at its output a spatiotemporal speckle pattern. To fully utilize the potential of MMFs, it is desired to have coherent control over the spectral and spatial degrees of freedom of the light that is transferred by the fiber.

The traditional way to unscramble information transferred by MMFs is using spatial light modulators (SLM) that shape the wavefront at the input and\slash or output of the fiber \cite{vcivzmar2012exploiting}. However, due to the wavelength-dependent transmission through multimode fibers, SLMs do not allow efficient control over a wide spectral bandwidth. Surprisingly, however, modal interference and modal dispersion in multimode fibers can in fact be harnessed for a wide range of applications \cite{cao2022harnessing}, such as fiber spectrometers \cite{redding2012using,redding2014high}, cryptography \cite{bromberg2019remote,amitonova2020quantum}, and optical implementations of neural networks \cite{teugin2021scalable}. Another challenge of multimode fibers is their extreme sensitivity to mechanical perturbations. We have recently utilized this sensitivity to make an all-fiber SLM, by applying controlled deformations to the fiber \cite{resisi2020wavefront}. Controlling the medium that the light propagates through, rather than the wavefront of the incident field, opens new opportunities for wavefront shaping, as was recently demonstrated in the microwave regime \cite{del2019optimally,del2021coherent}. These works highlight the great potential of controlling the medium instead of the incident field for extending wavefront shaping to the spectral domain, thanks to the strong coupling between the spectral and spatial degrees of freedom in complex media \cite{mccabe2011spatio,katz2011focusing}. 

In this work, we apply computer-controlled deformations to MMFs to realize all-fiber, reconfigurable spectral filters. To this end, we apply local bends at multiple positions along the fiber. The bends change the local propagation constants of the fiber's guided modes and induce mode mixing, thus changing the fiber’s wavelength-dependent transmission matrix. The curvatures of the bends are determined by an optimization algorithm that finds the optimal configuration that maximizes the overlap between the output spectrum and a target spectrum. Using this configuration, which we call fiber piano, we realize reconfigurable spectral filters with a 5pm resolution. 

\section{Experimental results}

\begin{figure}[!tbh]
\begin{centering}
\includegraphics[width=\columnwidth]{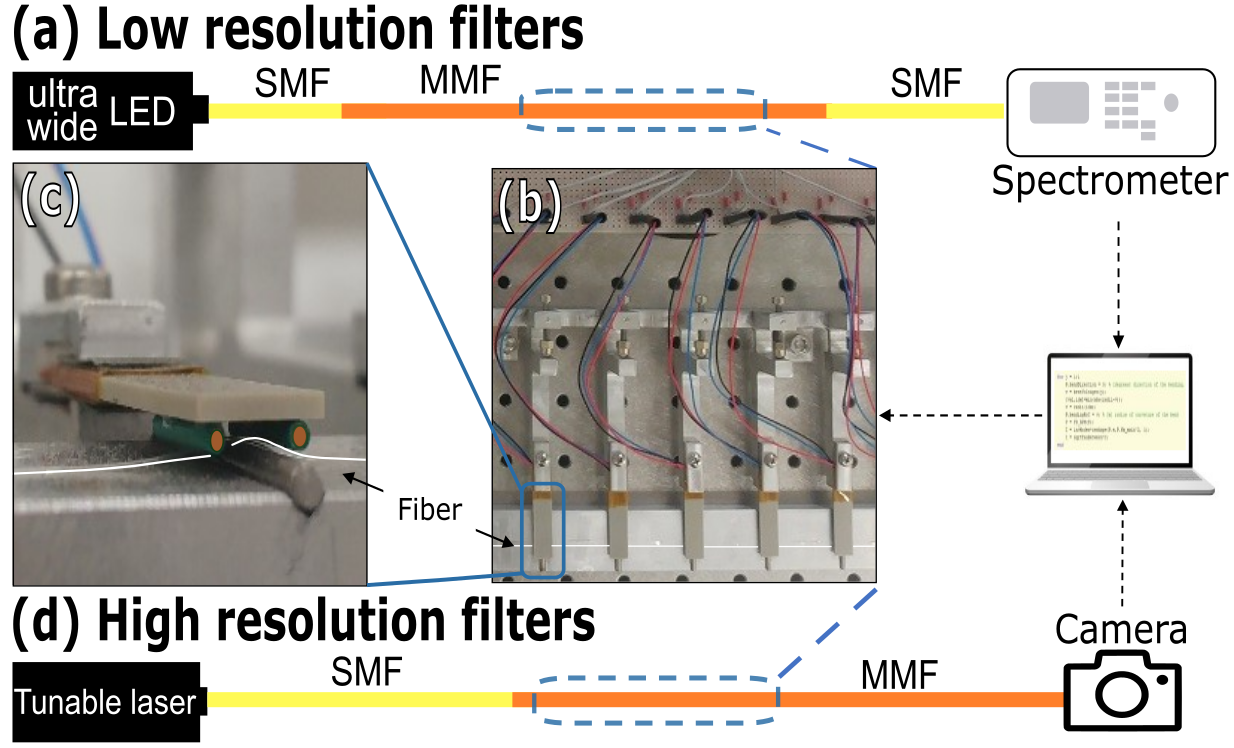}

\par\end{centering}
\caption{\textbf{Experimental setup for all-fiber spectral shaping.} (a) Light from a broadband LED source (Thorlabs MWWHF2) that is coupled to a single mode fiber (SMF), is injected into a 1m-long multimode fiber (MMF, Thorlabs FG050LGA). The output light is projected onto a single transverse mode by splicing the distal end of the MMF to a SMF (Thorlabs P1-630A-FC-1) that delivers the collected light to a spectrometer (Teledyne IsoPlane-320 and ProEm-HS:1024). (b) To control the transmission through the fiber we place an array of 28 computer-controlled piezoelectric actuators (both CTS ceramic NAC2225-A01 and Thorlabs PB4NB2W) along the fiber. Each actuator can travel $\sim$0.5mm in the vertical direction, pressing the fiber from above (c). The actuators create local deformations that change the propagation constants of the fiber's guided modes. An optimization algorithm (simulated annealing) searches for the optimal travel of each actuator, that maximizes the overlap of the output spectrum with the target spectrum. (d) To demonstrate spectral shaping with resolution beyond the spectral resolution of the spectrometer, we replaced the LED source with a fiber-coupled tunable laser at the telecom band (Keysight 81490A). The output facet for the MMF is imaged by an InGaAs camera (Xenics cheetah CL-7207) which records the wavelength-dependent speckle patterns.
}
\label{fig:Experimental Setup}
\end{figure}

The experimental setup is depicted in Fig.\ref{fig:Experimental Setup}a. A broadband fiber-coupled LED source is coupled to a single mode fiber (SMF), to obtain spatially coherent broadband light. The light is then coupled to a 1m-long step-index MMF, with a numerical aperture (NA) of 0.22 and a core diameter of 50$\mu$m. The output facet of the MMF is spliced to another SMF that is coupled to a spectrometer. The measured spectrum is determined by the complex wavelength-dependent interference of the fiber modes at the splicing point. We apply computer-controlled deformations to the fiber by placing 28 piezoelectric actuators along the fiber, with a 1.5cm separation between adjacent actuators (Fig.\ref{fig:Experimental Setup}b,c). We limit the minimal radius of curvature the actuators apply to 1cm, to avoid significant bending loss. As different deformations of the fiber yield different spectra at the output of the SMF, we search for the bend configuration that yields the required spectrum, using a simulated annealing optimization algorithm \cite{rutenbar1989simulated}.

\begin{figure}[!tbh]
\begin{centering}
\includegraphics[width=\columnwidth]{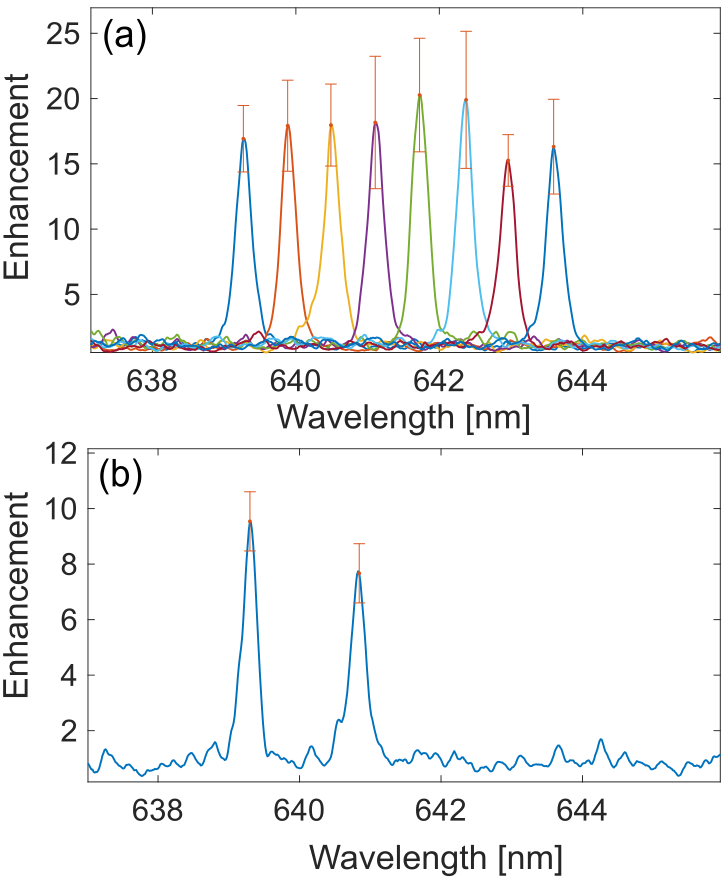}
\par\end{centering}
\caption{\textbf{Experimental realization of a tunable bandpass filter and a dual-band filter.} (a) Measured transmission spectra of the all-fiber tunable bandpass filter. Each curve represents an average of 6 spectra that were optimized to maximize the enhancement at the desired wavelength. We achieve a full-width-half-maximum (FHWM) of 245pm and enhancements in the range of 13 to 30. (b) Optimized transmission for a dual filter, with a FWHM of 241pm and separation of 1.5nm, obtained by maximizing the sum of square roots of the intensities at the chosen wavelengths. For the dual-band filters, we obtain enhancements in the range of 7.5 to 10.5. 
}
\label{fig:Configuration 1}
\end{figure}

To demonstrate spectral shaping, we realize a tunable bandpass filter, by maximizing the output intensity at different spectral bands (Fig.\ref{fig:Configuration 1}a). The full-width-half maximum (FWHM) bandwidth of each band is 245pm, determined by the spectral bandwidth of the MMF \cite{redding2012using}. Defining the enhancement factor as the peak-to-background ratio in the obtained spectra, we achieve enhancements in the range of 13 to 30. The efficiency of the system, defined by the ratio of output and input intensities at the chosen spectral band, is 0.06. To demonstrate the versatility of the spectral shaping, we also implement a dual bandpass filter, with two peaks separated by 1.5 nm and a FWHM of 241pm for each peak (see Fig.\ref{fig:Configuration 1}b).

\begin{figure}[!tbh]
\begin{centering}
\includegraphics[width=\columnwidth]{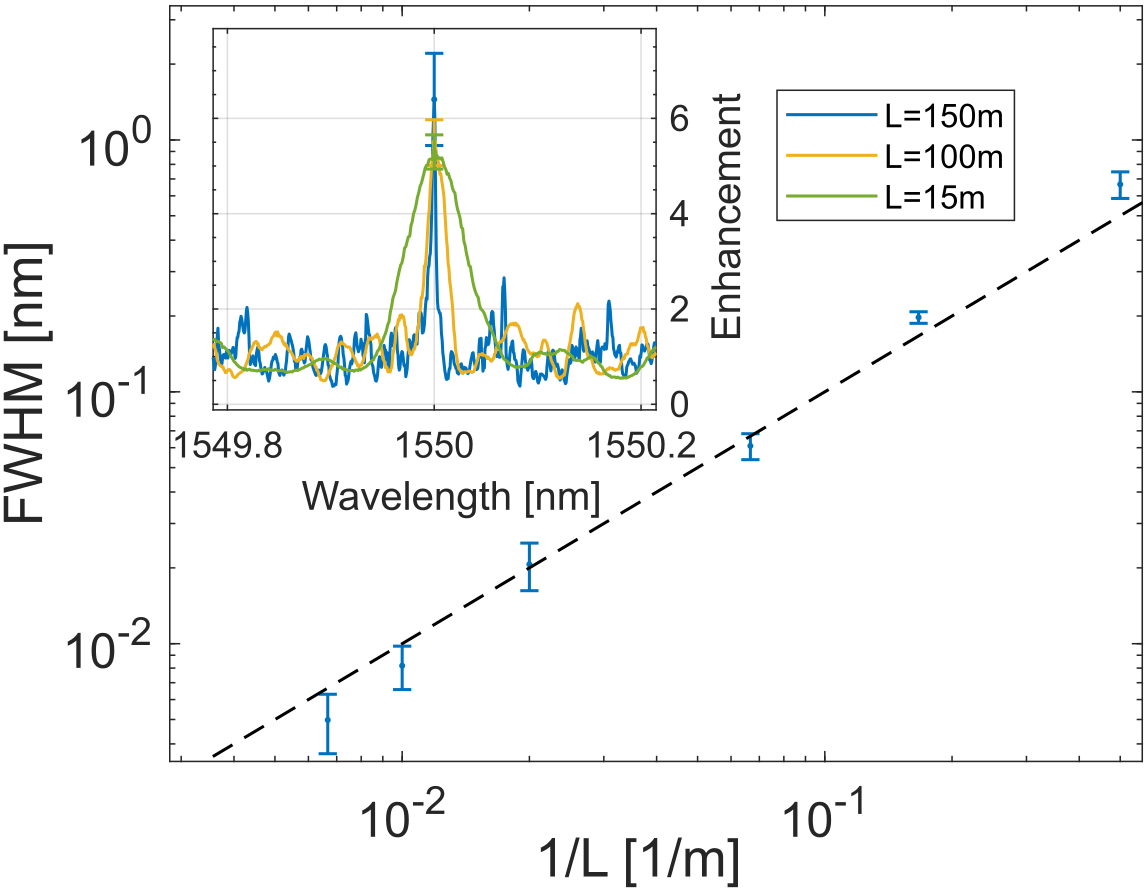}
\par\end{centering}
\caption{\textbf{Realization of narrow bandpass filters.} The measured FWHM of the optimized transmission peaks versus the fiber's length $L$ exhibits inverse linear scaling. The black curve is the predicted $1/L$ scaling \cite{redding2012using}. Inset: the transmission spectrum for $L=15$m (green), $L=50$m (yellow), and $L=150$m (blue). All spectra are normalized by the average background intensity. Errorbars in both graphs correspond to the standard deviation of different optimization runs.
}
\label{fig:configuration 2}
\end{figure}

Since the spectral bandwidth of the fiber is inversely proportional to the fiber's length \cite{redding2012using,redding2014high}, it is in principle possible to realize bandpass filters with extremely narrow linewidths by using long multimode fibers. In practice, however, the linewidths we could achieve were limited by the 60 pm resolution of the spectrometer used in the optimization process. We therefore replaced the spectrometer and the LED source with a tunable laser at telecom wavelengths, that can tune the wavelength in the C-band with a spectral resolution of 1pm, and an InGaAs camera that images the output facet of the MMF (Fig.\ref{fig:Experimental Setup}d). We fix the laser at a specific wavelength and apply the same optimization process described above, to enhance the intensity at a chosen square region of interest (ROI) on the camera. We set the width of the ROI to be equal to the FWHM of the intensity autocorrelation of the speckle pattern recorded by the camera. After the optimization, we scan the wavelength of the laser and observe the transmitted spectrum. The FWHM of the measured transmission peaks is inversely proportional to the fiber's length.  Using a 150m long fiber, we realize bandpass filters with a FWHM of 5 pm, and enhancement factors in the range of 5 to 7 (inset of Fig.\ref{fig:configuration 2})

\subsection{Numerical simulations}

To compare the performance of our system with its optimal limit, we numerically simulate the operation of the fiber-piano by computing the transmission spectra through a MMF with controlled local macro bends. We model the fiber-piano by numerically propagating scalar waves through curved segments of a MMF, separated by straight segments. Each curved segment simulates a macro bend induced by an actuator, where the travel of the actuator corresponds to a change in the radius of curvature of the segment. We use 44 discrete values for the radii of curvatures in the range of 1cm to 8cm, which roughly corresponds to the range of curvatures induced by the actuators in the experiment. 

To find the field at the output of the fiber, we calculate the transmission matrix (TM) of each segment and propagate the input field by multiplying it by the TM's of all the segments. To calculate the TM of the curved segments, we use "BPM-matlab" \cite{veettikazhy2021bpm}, an open source optical propagation simulation in MATLAB, which implements a finite difference beam propagation method for monochromatic light. We first compute the TM of the 44 curved segments to be used in the simulation, by propagating each guided mode of the fiber and projecting the output field on the guided mode basis. Propagation in the straight fiber segment is modeled by a diagonal TM, whose elements, that represent the phase accumulated by each guided mode, are computed by solving the Helmholtz equation for the fiber's index profile. The goal of the simulation is to find the radius of curvature of each bend, out of the 44 possible radii, that focuses the light at the distal end of the MMF, at a chosen wavelength. To this end, we use the same SA optimization algorithm used in the experiment. Once the optimal radii of curvature are found, we obtain the output spectrum by simulating propagation through the same configuration of bends and the same input field,  at different wavelengths.

\begin{figure}
\begin{centering}
\includegraphics[width=\columnwidth]{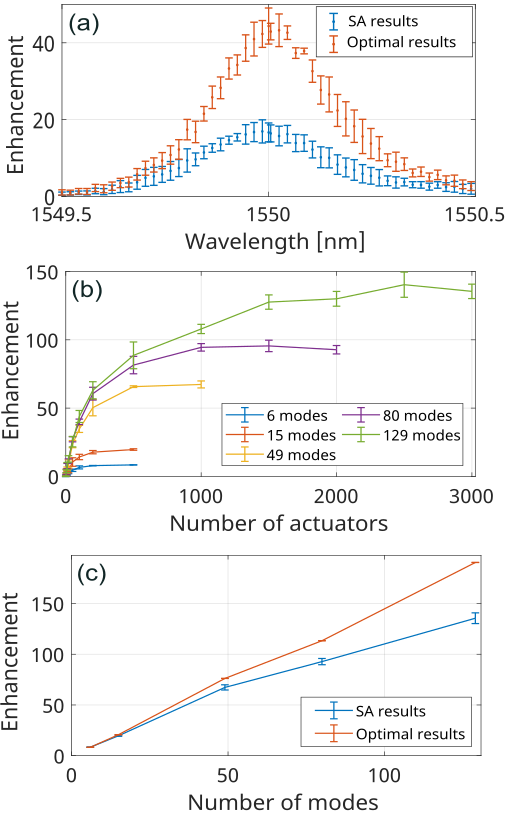}
\par\end{centering}
\caption{\textbf{Numerical simulation results.} (a) Simulation of the system with a 2m long fiber. We first run the SA optimization algorithm to search for the radii of curvature of the curved segments that maximize the intensity in the region of interest (ROI), at wavelength $\lambda=1550$ nm. Then for the same bend configuration and the same input modes, we propagate the field at different wavelengths and obtain the wavelength-dependent intensity in the ROI. For each wavelength, we present the enhancement, defined by the spatial peak-to-background ratio. The blue curve corresponds to the enhancements found by the optimization algorithm, while the red curve depicts the predicted enhancement for a perfect wavefront control (see text). (b) Simulated enhancements for different number of modes and actuators, showing that the enhancement increases with each of these parameters until saturation is observed. (c) The enhancement is achieved by the optimization algorithm (blue) and the optimal enhancements for perfect wavefront control (red), as a function of the number of guided modes in the fiber. The simulation data points are taken for the highest number of actuators that we tested for each fiber.}
\label{fig:simulation}
\vspace{-26.70367pt}
\end{figure}

We start by simulating propagation through a fiber-piano with the same number of actuators as in our experiments, 28 curved segments separated by straight segments. The fiber we simulate is a 2m-long step-index MMF, with a numerical aperture of 0.22 and a core diameter of 50$\mu$m. We excite the 50 first modes of the fiber, corresponding to the roughly 50 modes we excite in the experiment (see Fig.\ref{fig:MMF output}), with equal amplitudes and random relative phases. We achieve an enhancement of 15 and a FWHM of 350pm (blue curve in Fig.\ref{fig:simulation}a). Since we simulate the propagation over a bandwidth that is only slightly wider than the bandwidth of the fiber, here we define the enhancement as the spatial peak-to-background, instead of the spectral peak-to-background measured in the experiments. 
We attribute the difference from the experimental values obtained for the same fiber and the same wavelength range (enhancement of 7 and FWHM of 650pm)  to mode mixing that may increase the fiber's spectral bandwidth \cite{ho2011statistics} and system instabilities that may decrease the obtained peak-to-background ratios \cite{vellekoop2008phase}. 

It is instructive to compare the enhancement obtained by the fiber piano to the optimal enhancement one expects to achieve, for example using a perfect spatial light modulator at the proximal end of the fiber. We therefore look for the input field that maximizes the peak-to-background ratio using the same bend configuration that was found by the optimization algorithm for some arbitrary input. To this end, we first calculate the TM of the bent fiber as described above. Next, by multiplying the Hermitian conjugate of the TM by the target pattern (a focused spot in the desired ROI), we get the input field that maximizes the intensity in the ROI \cite{vellekoop2008phase}. Then, to find the output spectrum in the ROI, we numerically propagate this input field at different wavelengths. The optimal input field yields an enhancement of 45 and a FWHM of 320pm (red curve in Fig.\ref{fig:simulation}a). We therefore conclude that with 28 actuators we could achieve $\approx 30\%$ of the optimal enhancement.

To further explore the performance of the fiber-piano under optimal conditions, we simulate the achievable enhancement in a loss-free system, for an increasing number of fiber modes and actuators. Intuitively, the enhancement should rise as these two parameters are increased since they provide more degrees of freedom for the optimization process. The number of curved segments in the simulation is set by the number of actuators we simulate. The number of guided modes in the fiber is set by changing the core diameter while keeping the numerical aperture of the fiber fixed. We fix the total fiber length to 20m, and excite all the guided modes of the fiber with equal amplitudes and random relative phases. To neglect loss in the simulation, we force the transmission matrix of each segment to be unitary.

In Fig.\ref{fig:simulation}b, we depict the enhancement versus the number of actuators, for fibers with different number of modes. As expected, the enhancement increases with the number of actuators (Fig.\ref{fig:simulation}b). The saturated enhanced values obtained by the SA optimization, are close to the values expected by an optimal wavefront control, as depicted in Fig.\ref{fig:simulation}c, confirming that with enough actuators the fiber-piano can approach the optimal values. However, saturation is expected when the number of degrees of control, the actuators, approaches the number of guided modes of the fiber. Here we see the saturation is obtained at a much higher number of actuators, which indicates that the degrees of control in the simulation are not independent. Noticeably, as the number of modes grows the simulation performance deteriorates compared with the theoretical limit. This difference is most probably due to the increased probability that the SA algorithm finds a local minimum when the dimension of the optimization problem increases. 

\section{Conclusions and Outlook}

We experimentally realized all-fiber spectral shaping by applying computer-controlled deformations to a multimode fiber. As a proof of concept of this approach, we demonstrated a tunable bandpass filter with FWHM of 245pm, and a dual-wavelength filter with the same spectral width. Next, we showed the possibility to achieve narrower bandpass filters, using longer MMF, and achieved FWHM bandwidths as low as 5pm. Finally, we ran numerical simulations to test the performance of the fiber under optimal conditions. We show that with the fiber-piano it is possible to achieve peak-to-background ratios that are above $70\%$ of the ratios obtained for optimal wavefront control.   

In the high resolution configuration, where a camera that images the distal end of the fiber is used to select one spatial channel, one can simplify the detection method by replacing the camera with a SMF that is spliced at the distal end of the MMF along with a telecom photodiode. Our method demonstrates a first step towards realizing an all-fiber modulator for tailoring the spatial-spectral waveform at the output of multimode fibers, a key ingredient in a wide range of applications that are based on multimode fibers.

\section*{Funding Information}

This research was supported by the Zuckerman STEM Leadership Program and the ISF-NRF Singapore joint research program (grant No. 3538/20).


\section*{Disclosures}
The authors declare no conflicts of interest.

\printbibliography

\renewcommand{\thefigure}{S\arabic{figure}}
\setcounter{figure}{0}

\newpage
\section{Supplementary Material}

\subsection{Optimization Algorithm}

The propagation of each wavelength in the fiber is determined by the curvature of the bends, set by the travel of each actuator. Therefore, the spectrum at each spot on the distal end of the MMF is determined by the travel of each actuator. The optimization algorithm searches for the travels that maximize the intensity in a desired wavelength band (in the low resolution configuration) or in a desired ROI on the camera (in the high resolution configuration). 

Since the propagation of the light in each curved segment depends on the deformations induced by all the actuators prior to it, the ideal way to deal with such an optimization problem is using a nonlinear optimization algorithm. Here we use the simulated annealing (SA) algorithm \cite{rutenbar1989simulated},  which we found to yield the best results in terms of both the cost values and convergence times, which is about 10 minutes. 

The optimization parameters we used are initial temperature T of 20, with a cooling function that scales as T$\alpha^i$, where $i$ is the iteration number and $\alpha$ is 0.97 (0.99) in the experiments (simulations). The cost function maximizes the intensity in the spectral (spatial) window in the spectrum (image), normalized by the total intensity. We run the algorithm until we observe a sequence of 30 iterations without any improvement in the cost function. To decrease the probability of converging to local minima, at each iteration we add with some probability $p$ a weak perturbation to the optimal configuration of actuators and increase $p$ as the temperature decreases.

\subsection{System response time}

\begin{figure}[!tbh]
\begin{centering}
\includegraphics[width=\linewidth]{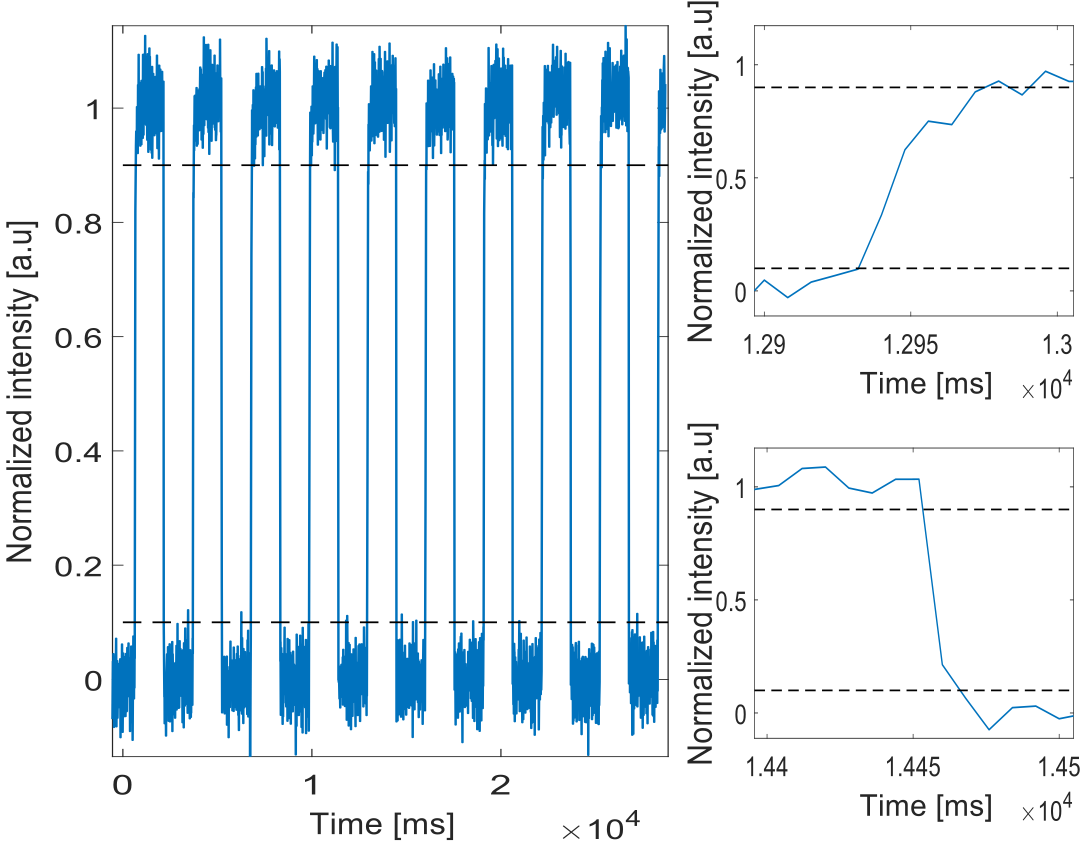}
\par\end{centering}
\caption{\textbf{System response time.}
Normalized intensity obtained when switching between the optimized configuration of the actuators and a random configuration. The rise time (10-90 percent from the optimal intensity) is 45ms (top right inset), and the fall time is 10ms (bottom right inset).}
\label{fig:switching}
\end{figure}

To measure the response time of the fiber-piano, we use the low resolution configuration and optimize the system for a specific spectral window. Then we switch between the optimal configuration of actuators and some other configuration. The response time of the system is measured by the time it takes to reach from 10 to 90 percent of the maximal intensity. We find that the rise time is 45ms, whereas the fall time is 10ms, as shown in Fig.\ref{fig:switching}. 

\subsection{System stability}

\begin{figure}[!tbh]
\begin{centering}
\includegraphics[width=\linewidth]{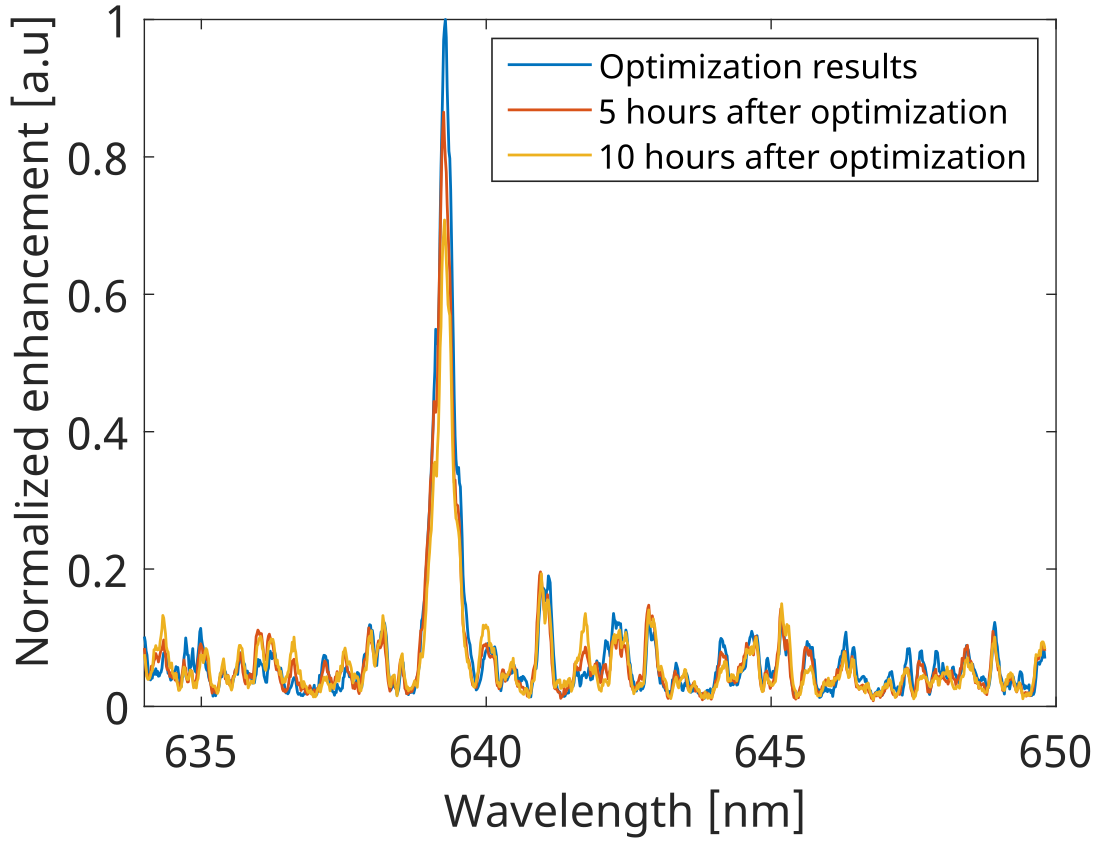}
\par\end{centering}
\caption{\textbf{System stability.} The measured spectra right after the optimization (blue curve), 5 hours after optimization (red curve), and 10 hours after the optimization (yellow curve). The peak intensity dropped by $\approx$$30\%$ after 10 hours.}
\label{fig:stability}
\end{figure}

To test the stability of the system, once we find the optimal configuration, we keep the actuators in the same position and record the output spectrum over time. We find that the spectrum maintains its shape and the overall transmissions fall by approximately $30\%$ after 10 hours (see Fig.\ref{fig:stability}).

\subsection{Simulation}
\subsubsection{Transmission matrix computation}
\begin{figure}[!tbh]
\begin{centering}
\includegraphics[width=\linewidth]{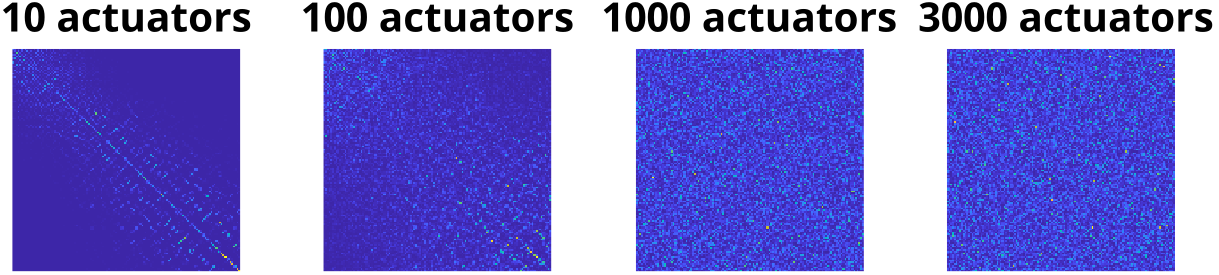}
\par\end{centering}
\caption{\textbf{Transmission matrix of the fiber for an increasing number of actuators $M$.} Each image represents the absolute square of the TM of the fiber, supporting 129 modes} 
\label{fig:TM}
\end{figure}

To simulate the operation on the fiber-piano, we chose 44 values for the radii of curvature, which model the bends induced by the piezoelectric actuators. The inverse of the radii of curvature are equally spaced between the maximal and minimal values. We compute the 44 transmission matrices (TM), ${C_1,...C_{44}}$, that correspond to the 44 values of the radii of curvature, by propagating the fiber's guided modes in the curved segment and projecting the output fields on the fiber mode basis. We then compute the TM of the fiber, which is deformed by $M$ actuators, by alternatively multiplying each one of the $M$ transmission matrices of the $M$ curved segments, $T_i\in{C_1,..C_{44}} ; i\in {1,..,M}$, with the diagonal matrix $D$: $T_{fiber}=DT_MD...T_2DT_1D$. $D$ represents the phase accumulated by each guided mode in a straight fiber. Fig.\ref{fig:TM} depicts the fiber's TM for an increasing number of actuators $M$, showing that as $M$ increases the TM evolves from a band-diagonal matrix to a fully randomized matrix. 

\subsubsection{Evaluation of the number of excited modes}
\begin{figure}[!tbh]
\begin{centering}
\includegraphics[width=\linewidth]{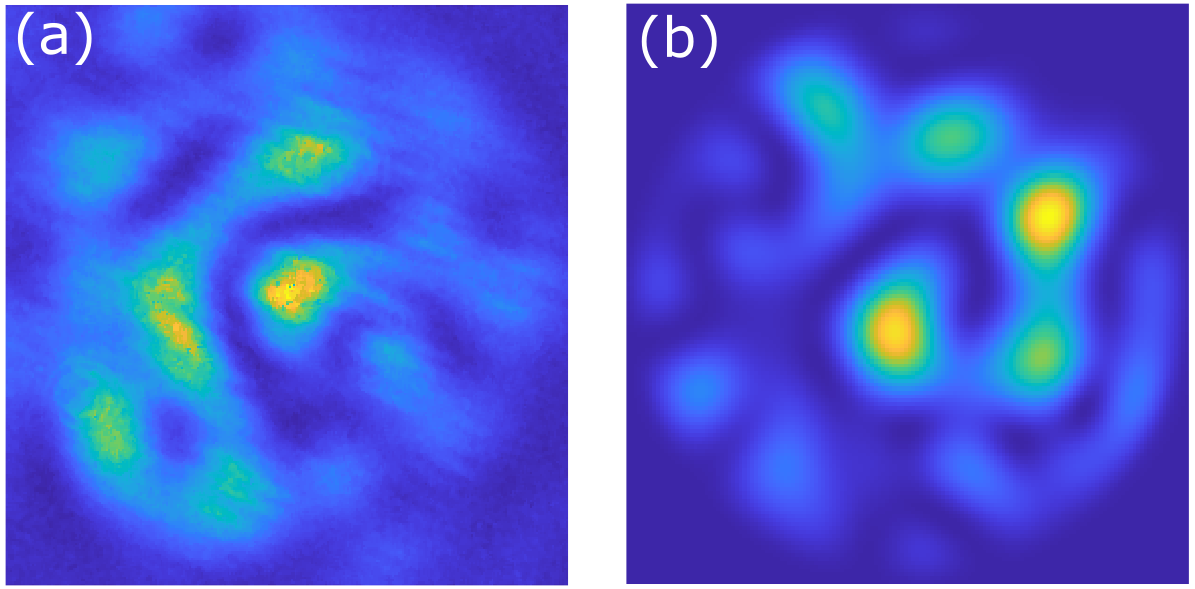}
\par\end{centering}
\caption{\textbf{Estimation of the number of excited modes.}
(a) Experimental and (b) simulated output of the MMF. In the simulation we found that we need to excited 50 modes in order to obtain roughly the same number of speckle grains and the same grain size as in the experiment.}
\label{fig:MMF output}
\end{figure}

For the simulations that model the experimental configuration with 28 actuators, we excite at the input only the first 50 modes, and not all the modes supported by the fiber, since by inspection of the speckle pattern obtained experimentally, we conclude that in our experiments we excite roughly 50 modes. The estimation of the number of excited modes is based on measuring the number of speckle grains at the distal end of the fiber, defined by the ratio of the core area and the speckle correlation area. Indeed, the experimentally measured speckle pattern has a similar grain size and number of grains to the speckle pattern simulated assuming excitation of the first 50 guided modes Fig.\ref{fig:MMF output}.

 \end{document}